\documentclass[12pt]{article}

\usepackage{latexsym}
\usepackage{amssymb}
\newcommand{\mysection}[1]{\section{#1}
   \hspace{0.8cm}\setcounter{equation}{0}}
\def\thefootnote{\fnsymbol{footnote}}
\def\ep{\epsilon}
\newcommand{\gappeq}{\mathrel{\rlap {\raise.5ex\hbox{$>$}}
    {\lower.5ex\hbox{$\sim$}}}}
\newcommand{\lappeq}{\mathrel{\rlap{\raise.5ex\hbox{$<$}}
    {\lower.5ex\hbox{$\sim$}}}} \newcommand{\myvev}[1]{{\left\langle #1
    \right\rangle}} 
\newcommand{\npb}{Nucl.\ Phys.}

\DeclareMathAlphabet   {\mathsc}{OT1}{cmr}{m}{sc}

\newcommand{\myref}[1]{(\ref{#1})}
\newcommand{\mysec}[1]{Section~\ref{#1}}

\newcommand{\pl}{{\it Phys.\ Lett. }}
\newcommand{\np}{{\it Nucl.\ Phys. }}

\def\Del{\Delta}
\def\half{{1\over2}}
\def\bea{\begin{eqnarray}}
\def\eea{\end{eqnarray}}
\def\beq{\begin{equation}}
\def\eeq{\end{equation}}

\def\bL{\bar{\Lambda}}
\def\ux{$U(1)_X$}
\def\uk{$U(1)_K$}
\def\uo{$U(1)$}
\def\dx{\delta_X}
\def\vx{V_X}

\def\superint{\int d^{4}\theta}
\def\M{\bar{M}}

\newcommand{\WaWa}{ W_a^{\alpha} W^a_{\alpha}}

\newcommand{\Wa}{W_{\alpha}}

\newcommand{\Wc}{W^{\alpha}}

\def\re{{\rm Re}}
\def\im{{\rm Im}}

\def\W{\overline{W}}

\def\G{{\cal G}}

\def\tr{{\tilde r}}

\def\tX{{\widetilde X}}

\def\tF{{\tilde F}}

\def\D{{\cal D}}

\def\bD{\bar{\D}}

\def\pp{\partial}

\def\ibar{\bar{\imath}}

\def\[{\left [}
\def\]{\right ]}
\def\({\left (}
\def\){\right )}

\def\r{\right|}
\def\l{\left.}

\def\H{\bar{H}}

\def\T{\bar{T}}

\def\z{\bar{z}}
\def\q{\bar{q}}
\def\del{\delta}

\def\S{{\bar{S}}}

\def\Tr{{\rm Tr}}

\def\Ti^{T^{(i)}}

\def\f{\bar{f}}

\def\L{{\cal L}}

\def\n{\bar{n}}
\def\m{\bar{m}}

\def\Z{{\bar{Z}}}

\def\bph{\bar{\phi}}
\def\bPh{\bar{\Phi}}

\def\bF{\bar{F}}

\def\bj{\bar{\jmath}}
\def\a0{\alpha_0}
\def\chiproj{(\bD^2 - 8R)}
\def\bchiproj{(\D^2 - 8 \bar R)}

\def\eee{\nonumber \\ &=&}
\def\ddd{\nonumber \\ &&}
\def\mmm{\nonumber \\ &&\qquad}
\def\nnn{\nonumber \\ }
\def\hc{ + {\rm h.c.}}

\begin{document}

\begin{titlepage}
\begin{center}

\hfill February 2016 \\[.3in]

{\large {\bf Quantum supergravity, supergravity anomalies and string
    phenomenology}}\footnote{This work was supported in part by the
  Director, Office of Science, Office of High Energy and Nuclear
  Physics, Division of High Energy Physics, of the U.S. Department of
  Energy under Contract DE-AC02-05CH11231, in part by the National
  Science Foundation under grant PHY-1316783.}\footnote{To be
  published in a memorial volume for Raymond Stora.} \\[.2in]

 Mary K. Gaillard
\\[.1in]

{\em Department of Physics and Theoretical Physics Group,
 Lawrence Berkeley Laboratory, 
 University of California, Berkeley, California 94720}\\[.5in] 

\end{center}

\begin{abstract}
\noindent
I discuss the role of quantum effects in the phenomenology of
effective supergravity theories from compactification of the weakly
coupled heterotic string. An accurate incorporation of these effects
requires a regularization procedure that respects local supersymmetry
and BRST invariance and that retains information associated with the
cut-off scale, which has physical meaning in an effective
theory.  I briefly outline the Pauli-Villars regularization
procedure, describe some applications, and comment on what remains to
be done to fully define the effective quantum field theory.
\end{abstract}
\end{titlepage}

\newpage

\renewcommand{\thepage}{\roman{page}}
\setcounter{page}{2}
\mbox{ }

\vskip 1in

\begin{center}
{\bf Disclaimer}
\end{center}

\vskip .2in

\begin{scriptsize}
\begin{quotation}
This document was prepared as an account of work sponsored by the United
States Government. While this document is believed to contain correct 
 information, neither the United States Government nor any agency
thereof, nor The Regents of the University of California, nor any of their
employees, makes any warranty, express or implied, or assumes any legal
liability or responsibility for the accuracy, completeness, or usefulness
of any information, apparatus, product, or process disclosed, or represents
that its use would not infringe privately owned rights.  Reference herein
to any specific commercial products process, or service by its trade name,
trademark, manufacturer, or otherwise, does not necessarily constitute or
imply its endorsement, recommendation, or favoring by the United States
Government or any agency thereof, or The Regents of the University of
California.  The views and opinions of authors expressed herein do not
necessarily state or reflect those of the United States Government or any
agency thereof, or The Regents of the University of California.
\end{quotation}
\end{scriptsize}

\vskip 2in

\begin{center}
\begin{small}
{\it Lawrence Berkeley Laboratory is an equal opportunity employer.}
\end{small}
\end{center}

\newpage

\def\thefootnote{\arabic{footnote}}
\renewcommand{\theequation}{\arabic{section}.\arabic{equation}}
\renewcommand{\thepage}{\arabic{page}}
\setcounter{page}{1}
\def\thefootnote{\arabic{footnote}}
\setcounter{footnote}{0}

\mysection{Introduction}
Since the first ``string revolution'' of 1984, starting with the
Green-Schwarz discovery~\cite{gs} that string theories with an
$SO(32)$ or $E_8\otimes E_8$ gauge sector are anomaly-free, there has
been a considerable amount of work on orbifold compactifications of
the heterotic $E_8\otimes E_8$ string theory~\cite{wchs} that mimic
Calabi-Yau compactification~\cite{cy} on a six-dimensional manifold.
In these studies, the favored mechanism for supersymmetry breaking has
been gaugino condensation in the subgroup of the hidden sector $E_8$
that survives after symmetry breaking by Wilson lines, aka the
Hositani mechanism.  Gaugino condensation is an inherently quantum
effect in the effective supergravity theory that is the large tension
limit of the string theory, and, as described below, quantum anomalies
play an essential role in its description.  Other aspects where
quantum effects play a role include the issues of vacuum stability and
flavor changing neutral currents, axion physics and soft supersymmetry
breaking. The last of these have contributions that are
specific to supergravity, and a reliable method for computing
them is imperative.

Specifically, we require a regularization procedure that respects
gauge invariance and local supersymmetry.  In renormalizable, globally
supersymmetric theories one uses dimensional reduction.  Like
dimensional regularization, used for ordinary gauge theories, this
procedure eliminates quadratic divergences altogether.  However, in
an effective theory, such as a four-dimensional supergravity theory
from a ten-dimensional string theory, the quadratic divergences, or more
specifically the effective cut-offs, have physical significance.  For
example, the effective cut-off of a few hundred GeV for the Fermi
theory of weak interactions pointed the way towards relevant energies
to search for new physics, and this new physics manifested itself in
the form of the $W$ and $Z$ bosons with masses of about 100 GeV.
Similarly, the need to suppress strangeness-changing neutral currents
indicated a scale of a few GeV, and led to the successful prediction
of the charmed quark mass.  In a simple field theory one can just
introduce a momentum cut-off, which generally gives correct results at
one-loop, up to the precise coefficient of the quadratically divergent
operators.  However this procedure does not respect local symmetries,
or even global supersymmetry, which is why one uses dimensional
regularization or reduction in renormalizable theories with these
symmetries. In the case of local supersymmetry, or supergravity, the
use of a momentum cut-off can produce misleading results, as
illustrated in some examples below.

The ultraviolet divergent part of the on-shell effective Lagrangian
for a general supergravity theory with at most two derivatives at
tree-level was determined~\cite{us} using the covariant derivative
expansion~\cite{mkg}. As is well known, the quadratically divergent
contribution is prescription-dependent.  Specifically, the use of a
simple cut-off or subtraction procedure for a supersymmetric theory
does not yield a supersymmetric result.  However, as first shown in
~\cite{casimir}, when the theory is regulated by Pauli-Villars fields
embedded in a supersymmetric Lagrangian, there are additional finite
terms quadratic in the PV masses that complete the one-loop action in
such a way that the result is supersymmetric.

\mysection{Pauli-Villars regularization of supersymmetry and
  supergravity}\label{PV} 
Pauli-Villars (PV) regularization was
initially used to regulate the divergences in quantum electrodynamics.
However this procedure could not be generalized to non-Abelian gauge
theories because the introduction of the needed massive PV gauge
bosons breaks gauge invariance, or--in the gauge-fixed version of the
theory--BRST invariance~\cite{brst}.  However, in supersymmetric
theories, the same cancellation of ultraviolet divergences that leads
to the well-known nonrenormalization theorems also allows for PV
regularization of these theories.

For example, some of the divergences arising from loop diagrams
involving gauge boson self-couplings, that require the introduction of
massive vector PV fields, are canceled in supersymmetric theories by
loops involving gauginos.  As a consequence a renormalizable
supersymmetric theory can be regulated by introducing PV fields only
in chiral supermultiplets, and BRST invariance is unbroken.  In the
case of supergravity, the theory can be regulated by the introduction
of massive PV chiral supermultiplets and Abelian vector multiplets,
and BRST invariance again remains unbroken.  As a result, all the
on-shell logarithmic and quadratic divergences can be
canceled~\cite{pv} in a supergravity theory defined in the usual
way~\cite{sugra} by an arbitrary holomorphic superpotential $W(Z)$, a
real K\"ahler potential $K(Z,\Z)$ and a holomorphic gauge kinetic
function $f(Z)$, provided the gauge charges of matter fields have the
same overall quadratic Casimirs~\cite{bg}
\noindent
as those of some real (reducible) representation $R$ of the Yang-Mills
gauge group:
\noindent

\beq C^a_M = \Tr T^a T^a \equiv C^a_R\label{cascond}.\eeq
\noindent
Here $Z$ represents the chiral matter superfields, and the sum in the
trace \myref{cascond} runs over all the chiral fields $Z^i$.  The
condition \myref{cascond} on the Casimirs is satisfied in the Minimal
Supersymmetric Standard Model (MSSM) and in extensions thereof.  In
addition to two Higgs doublets, the MSSM, has $2N_f$ fundamental
representations (reps) ${\bf n}$ of each group factor $\G_n=SU(n),\;
n=2,3$, where $N_f$ is the number of quark flavors. Their Casimirs can
be mimicked by $N_f$ real PV reps $({\bf n} + {\bf\n})$. Further
extensions necessarily involve real representations of the Standard
Model gauge group, so that the additional states can get SM gauge
invariant masses.  The condition (\ref{cascond}) is also satisfied in
the hidden sectors~\cite{joel} that can accompany the
Standard-Model-like $\mathbf{\mathbb{Z}}_N$ orbifolds found
in~\cite{iban}.  These hidden sectors also come in even numbers of
representations, except for two cases. In one the hidden sector
contains 3 {\bf 16}'s of $SO(10)$ which contribute $C^{SO(10)}_M = 6$;
this can be mimicked by a real PV rep with 6 {\bf 10}'s. The other has
a hidden sector with 3 $({\bf 5} + {\bf 10})$'s and 6 ${\bf\bar 5}$'s
of $SU(5)$ with $C^{SU(5)}_M = 9$, that can be mimicked by 9 real PV
reps $({\bf 5} + {\bf\bar5})$.  Since the underlying theory is finite
when all degrees of freedom are included, one would expect
(\ref{cascond}) to have a solution for general superstring
compactifications.

The part of the resulting one-loop action that is quadratic in the PV
masses is just a renormalization of the K\"ahler potential, while the
part logarithmic in PV masses contributes to the renormalization of
both the K\"ahler potential and the gauge kinetic function (which is
no longer holomorphic at the quantum level).  In addition there are
new operators of dimension 6-12; those of dimension six involve the
curvature of the K\"ahler metric and derivatives of the gauge kinetic
function.

Most of the linear divergences of a generic supergravity theory can be
canceled by the PV fields introduced above. Their associated chiral
anomalies\footnote{The chiral anomalies of supergravity were first
  evaluated in~\cite{danf}, including those arising from an additional
  connection~\cite{bdkv} in theories with an anomalous \uo\, and no
  compensating Green-Schwarz term; these contributions are not present
  in the class of string-derived theories considered here.}  either
disappear or reappear through noninvariant PV mass terms, forming an
``F-term'' anomaly that incorporates the associated conformal
anomaly~\cite{bg}.  However there are chiral anomalies associated with
the affine connection in the gravitino covariant derivative and with
an off-diagonal gravitino-gaugino connection that cannot be canceled
by the PV fields.  These form supersymmetric ``F-term'' anomalies
together with conformal anomalies associated with total
derivatives that are not canceled by the PV fields, provided the
cut-off is field dependent:
\noindent
\beq \Lambda(Z,\Z) = \mu_0\exp(K/4),\label{lambdaz}\eeq
\noindent
where $\mu_0$ is a constant that can be set to infinity at the end of
the calculation; the only effect of the field-dependence in
\myref{lambdaz} is that total derivatives with nonvanishing
coefficients of $\ln\Lambda$ do not drop out of the S-matrix of the
regulated theory. For example, the conformal anomaly associated with
the Gauss-Bonnet term combines with the chiral anomaly proportional to
the space-time curvature term $r\cdot\tr$, with an overall
coefficient~\cite{bg} that agrees with string loop
calculations~\cite{agn}.

In addition to the above-described ``F-term'' anomalies, there
are ``D-term'' anomalies associated with logarithmic divergences
that are not canceled by the PV regulator fields, and that have
no chiral counter-parts.  

The anomalies that we are concerned with here involve
symmetries of the underlying string theory that are not respected at
the quantum level of the effective field theory without the
introduction of some cancellation mechanism that appears only at
one-loop order. In compactifications of the weakly coupled heterotic
string theory these are a discrete group of transformations known as
``T-duality'' or ``target space modular invariance''~\cite{mod},
present in all heterotic string compactifications, and an anomalous
\uo\ symmetry, often referred to as \ux, that is present in most
compactifications involving Wilson lines. Gauge symmetry breaking by
Wilson lines is generally needed both for providing a gauge group that
resembles the Standard Model and for gaugino condensation in the
hidden sector, which can provide a source of supersymmetry breaking.

The effective four dimensional (4d) theory from the heterotic string
includes several important ``moduli'' chiral supermultiplets: the
dilaton supermultiplet $S$, whose vacuum value determines the gauge
coupling constant and the $\theta$-parameter of the 4d gauge theory,
and ``K\"ahler moduli'' $T^i$ whose vacuum values determine the size
and shape of the compact six dimensional space. There are at least
three of the latter in orbifold compactifications, and the group of
T-duality transformations always contains an $SL(2,\bf Z)$ subgroup
under which these three ``diagonal moduli'' transform as
\noindent
\beq T'^i = {a T^i - i b\over i c T^i + d},\qquad a d - b c
=1, \label{Ttransf}\eeq
\noindent
and which is generated by two elements: the inversion of the radii
(in string units) $\re t^i \to 1/\re t^i$ of the three 2-tori
in the compact six dimensional space, and the axionic shifts 
$\im t^i \to \im t^i + 1$.  Here $t^i = \l T^i\r$ is the scalar
component of the chiral superfield $T^i$.  More generally, 
T-duality acts as follows on chiral (antichiral)
superfields $Z^p = T^i,\Phi^a\;(\Z^{\bar p} = \T^{\ibar},\bPh^{\bar a})$:
\beq T^i\to h(T^j), \qquad \Phi^a \to
f(q^a_i,T^j)\Phi^a,\qquad \T^{\ibar}\to h^*(\T^{\bj}), \qquad \bPh^{\bar a}\to
f^*(q^a_i,\T^{\bj})\bPh^{\bar a},\label{tduality}\eeq
where $q^a_i$ are the modular weights of $\Phi^a$, and,
under \ux\, transformations,
\beq\vx \to \vx + \Lambda_X + \bar\Lambda_X,\qquad
\Phi^a \to e^{-q^a_X\Lambda_X}\Phi^a,\qquad
\bar\Phi^a \to e^{-q^a_X\bL_X}\bar\Phi^a,\label{UX} \eeq
where $\vx$ is the \ux\, vector superfield, with $\Lambda_X\;
(\bar\Lambda_X)$ chiral (antichiral), and $q^a_X$ are \ux\, gauge
charges.

In order to faithfully represent the underlying string theory, in
which both of the above symmetries are exact to all orders in
perturbation theory~\cite{mod}, additional terms must be added to the
effective supergravity Lagrangian.  The terms that restore the
symmetry to the coefficients of bilinears in the Yang-Mills fields and
the space-time Riemann tensor at one loop were identified some time
ago as a combination of four-dimensional counterparts~\cite{gs4} of the
ten-dimensional Green-Schwarz term~\cite{gs} and, for some
compactifications, threshold corrections~\cite{th} that contribute to
the cancellation of the modular anomaly.  The implementation of these
cancellations is possible only if the loop corrections in the regulated
theory satisfy certain constraints.  For example, in the absence of
threshold corrections, the gauge charges and modular weights must
satisfy:
\noindent
\bea 8\pi^2b &=& {1\over24}\(2\sum_p q^p_i - N + N_G -
21\)\quad\forall\quad i\eee C^a - C_M^a + 2\sum_b(T^2_a)^b_b q^b_i
\quad \forall\quad i,a,\label{wthconds} \\
\noindent
2\pi^2\dx &=& - {1\over24}\Tr T_X = - {1\over3}\Tr T_X^3 = -
\Tr(T^2_aT_X)\quad \forall\quad a\ne X\label{uxconds},\eea
where $C^a$ is the quadratic Casimir in the adjoint of the gauge
group factor $\G_a$, and the matter Casimir $C^a_M$ is defined in
\myref{cascond}.

The above expressions for the coefficients of one-loop generated
operators that are linear in the parameters $q^p_i,\,q^a_X$ of
the anomalous transformations are universal.  They are independent of
the precise choice of PV regulator fields, provided the one-loop,
on-shell quadratic and logarithmic divergences are canceled.  However
this is not the case for operator coefficients that are quadratic and
higher order in these parameters~\cite{ssanom,bg}.  We will return to
this issue in \mysec{unfin}.

\mysection{Quadratic divergences}
It has been pointed out~\cite{ckn,clm} that the loop suppression
parameter
\beq \ep = {1\over16\pi^2}\eeq 
\noindent
may be compensated by large coefficients, leading to significant
effects from loop corrections. For example, if supersymmetry is F-term
dominated with negligible vacuum energy $\langle{V}\rangle \approx0$
at tree level, the quadratically divergent correction to the scalar
potential reduces to:
\beq V_Q = \ep\Lambda^2\[\(N_\chi - 1\)|M_\psi|^2 - N_G|M_\lambda|^2
- R_{i\m}F^i\bF^{\m}\],\label{vquad}\eeq
\noindent
where $N_\chi$ and $N_G$ are the number of chiral and gauge
supermultiplets, respectively, $M_\psi = e^K W$ is the gravitino
mass, $M_\lambda$ is the gaugino mass, which depends on derivatives
of the kinetic function $f(z) = \l f(Z)\r,\;z = \l Z\r,$ and
$R_{i\m}$ is the Ricci tensor associated with the K\"ahler metric
$K_{i\m}.$ 

\subsection{Vacuum stability} 
Typical orbifold compactifications have many
more chiral multiplets than gauge multiplets: $N_\chi\gappeq 300,\quad
N_G \lappeq 65$. In addition, in many gravity mediated
supersymmetry-breaking scenarios the gaugino mass $M_\lambda$ is much
smaller than the the gravitino mass:
\beq M_\lambda^2 = {1\over4}f_i\f^i M^2\ll M^2.\eeq
Thus the first term in \myref{vquad} suggests the possibility of a
significant {\it positive} contribution to the vacuum
energy~\cite{ckn}, perhaps curing the problems with classes of models
that have negative vacuum energy at tree level.  However, in the
regulated theory \myref{vquad} is replaced by
\beq  V_Q\to\ep\[|M|^2\(N_\chi\Lambda^2_\chi
- \Lambda^2_{\rm  grav}\) - N_G M^2_\lambda\Lambda^2_G
- R_{i\m}F^i\bF^{\m}\Lambda'^2_\chi\] + \cdots,\label{vreg}\eeq
\noindent
where the ellipsis indicates finite terms proportional to the PV
squared masses such that the one-loop quadratically divergent
corrections are completely absorbed into renormalizations:
\beq\L_Q = \L_{\rm tree}(g^R_{\mu\nu},K^R) - 
\L_{\rm tree}(g_{\mu\nu},K) + O(\ep^2),\qquad
K^R = K + \ep{\sum_A}\Lambda^2_A.\label{lquad}\eeq
\noindent
The effective squared cut-offs $\Lambda^2_A$ in \myref{vreg} and
\myref{lquad} are determined by combinations of the PV masses $M_I$ 
weighted by their signatures $\eta_I = \pm 1$:
\beq \Lambda^2_A = \sum_I C^I_A \eta_I M^2_I\ln M^2_I,\qquad
\sum_{I\in A}\eta_I M^2_I = 0,\eeq
\noindent 
where $C^I_A$ is a constant.  In fact, the apparent appearance of a
sizable, positive cosmological constant in \myref{vquad} and
\myref{vreg} is misleading on several counts~\cite{quaddiv}.  The sign
of $\Lambda^2_A$ is, in fact, indeterminate~\cite{sigma} if there are
five or more terms in the sum, which is generally required to
eliminate all the UV divergences of SUGRA. More importantly, if
$N_\chi\sim\ep^{-1}$ one has to sum the leading $(\ep\Lambda^2)^n$
terms, and supersymmetry dictates that the higher order terms complete
the Lagrangian $\L_{\rm tree}(g^R_{\mu\nu},K^R)$ with $K_R$ given by
\myref{lquad}.  So, for example, if the $M^2_I$ are field independent
constants, we just get for this contribution to the loop corrected
potential
\noindent
\bea V_Q &=& e^{K+\Del K}\[\(W_i + K_i\)K^{i\m}\(\W_{\m} + K_{\m}\)
- |W|^2\] + {\re f\over2}D^a D_a, \nnn W_i &=& {\pp\over\pp z^i}W = -
  e^{-K/2}K_{i\m}\bF^{\m},\eea
\noindent
where $K^{i\m}$ is the inverse of the K\"ahler metric, $F^m =
(\bF^{\m})^\dag$ is the auxiliary field for the chiral superfield
  $Z^m$, and
\noindent
\beq D_a = K_i(T_a z)^i = K_{\m}(T^T_a\z)^{\m},\qquad K_i = {\pp
  K\over\pp z_i}, \quad K_{\m} = {\pp K\over\pp z_{\m}},\eeq
\noindent
is the auxiliary field for the vector supermultiplet $V_a$.
If, in addition, supersymmetry is broken only by F-terms, $\myvev{D_a}
= 0$, the vacuum energy is just multiplied by a positive constant,
so if it vanishes at tree level, there is no large correction of 
order $\ep N$.

\subsection{Flavor changing neutral currents} \label{fcnc}
It was also pointed out~\cite{clm} that the last term in \myref{vquad}
or \myref{vreg} can be significant because the contracted indices of
the K\"ahler Riemann tensor implicit in the Ricci tensor imply a sum
over all the chiral supermultiplets.  The K\"ahler potential for the
twisted sector from orbifold compactification of the heterotic string
is not known beyond leading (quadratic) order, and could include terms
that induce flavor changing neutral current (FCNC) effects in the
observable sector.  Experimental limits on these effects therefore
imply restrictions on the tree-level K\"ahler potential.  A sufficient
condition~\cite{quaddiv} for a ``safe'' K\"ahler potential at the
quantum level is the presence of isometries of the K\"ahler geometry.
For example, the K\"ahler potential for an untwisted sector $(i)$ from
orbifold compactification takes the form
\noindent
\beq K^{(i)} = - \ln\(T^i + \bar T^{\bar i} -{\sum_{a=1}^{N_i}}
|\Phi^a_i|^2\),\label{noscale} \eeq
which has an $SU(N_i + 1,1)$ symmetry that is necessarily also 
a symmetry of the Ricci tensor:
\beq R^{(i)}_{p\q} = (N_i + 2)K^{(i)}_{p\q},\qquad p,q = i,a\in(i).\eeq
Alternatively the suppression of FCNC effects can by achieved
through a judicious choice of PV masses~\cite{quaddiv}.

\mysection{Anomaly cancellation and its implications} 
\noindent
In $\mathbf{\mathbb{Z}_3}$ and $\mathbf{\mathbb{Z}_7}$
compactifications of the heterotic string, with no threshold
corrections, the variation under \myref{tduality} and \myref{UX} of
the anomalous part of the one-loop corrected Lagrangian contains the term
\beq \Del\L_{\rm anom} = {1\over8}\int d^4\theta{E\over R}\Phi H\hc,
\label{DelL}\eeq
\noindent
as expressed in the K\"aher \uo\, [\uk] superspace formulation of
supergravity~\cite{bggm}.  Here $E$ is the superdeterminant of the
supervielbein and $R = \half e^{K/2}W,\;\l R\r = \half M_\psi$ is an
auxiliary field of the supergravity supermultiplet. Also
\noindent
\beq \Phi = {1\over3}W^{\alpha\beta\gamma}W_{\alpha\beta\gamma} +
W^\alpha_a W^a_\alpha\label{Phi} \eeq
\noindent
is a chiral superfield with \uk\, weight 2 with $W^a_\alpha$ and
$W^{\alpha\beta\gamma}$ the Yang-Mills and spacetime curvature
superfield strengths, respectively, and
\beq H = -b F(T) + \half\dx\Lambda_X \label{defH}\eeq
\noindent
is a zero weight chiral supermultiplet, with $b$ and $\dx$ subject
to the conditions \myref{wthconds} and \myref{uxconds}. The K\"ahler
potential can be decomposed as
\beq K = G(T,\T) + K_{\rm inv}\eeq
\noindent
with $K_{\rm inv}$ modular invariant, and
\beq G(T,\T) \to G(T,\T) + F(T) + \bF(\T) \eeq
under the T-duality transformation \myref{tduality}.  In component
notation, \myref{DelL} reads
\noindent
\bea \Del\L_{\rm anom} &=& - {1\over4}\sqrt{g}\[\re
H\(F_a^{\mu\nu}F^a_{\mu\nu} - {2\over x^2}\D_a\D^a\) + \im H
F_a\cdot\tF^a\]\ddd + {\sqrt{g}\over96}\[\re
H\(r^{\mu\nu\rho\sigma}r_{\mu\nu\rho\sigma} - 2r_{\mu\nu}r^{\mu\nu} +
{1\over3}r^2\) + \im H r\cdot\tr\] \nonumber \\ & & +
{\sqrt{g}\over144}\(\re H X_{\mu\nu}X^{\mu\nu} + \im
H\tX_{\mu\nu}X^{\mu\nu}\) + {\rm fermions},\label{d3}\eea
where $X_{\mu\nu}$ is the field strength associated with the
K\"ahler \uo\, connection in the fermion covariant derivatives.

\subsection{Anomaly cancellation}
Anomaly cancellation is most readily implemented using the linear
multiplet formulation for the dilaton~\cite{linear}.  A linear
supermultiplet is a real superfield that satisfies
\beq (\D^2 - 8\bar R)L = (\bar\D^2 - 8R)L = 0,\eeq
\noindent
where $\bar\D^2 - 8R$ is the chiral projection operator in
supergravity.  The superfield $L$ has three components: a scalar, the
dilaton $\ell = \l L\r$, a spin-$\half$ fermion, the dilatino $\chi$,
and a two-form $b_{\mu\nu}$ that is dual to the axion $\im s$; it has
no auxiliary field.  For the purpose of anomaly cancellation we want
instead to use a real superfield that satisfies the {\it modified}
linearity condition:
\beq (\bar\D^2 - 8R)L = -\Phi,\qquad (\D^2 - 8\bar R)L = 
- \bar\Phi,\label{lincond}\eeq 
\noindent
where $\Phi$ is a chiral multiplet with \uk\, and Weyl
weights~\cite{bggm} $w_K(\Phi) = 2,\; w_W(\Phi) = 1,$ respectively.
Consider a theory defined by the K\"ahler potential $K$ and
the kinetic Lagrangian $\L_{\rm K E}$:
\noindent
\beq K = k(L) + K(Z,\Z),\qquad \L_{\rm K E} =
-3\displaystyle{\superint}\,E\,F(Z,\Z,\vx,L)
\label{Llin1}.\eeq
\noindent
The condition for a canonical Einstein term in \uk\, superspace is
give by
\beq
F- L\frac{\partial F}{\partial L} = - L^2{\pp\over\pp L}\({1\over L}F\)
 = 1- \frac{1}{3}L \frac{\partial k}{\partial L},\label{eincond}\eeq
with the solution:
\beq F(Z,\Z,\vx,L) = 1 + \frac{1}{3} L V(Z,\Z,\vx) +
\frac{1}{3}L \int \frac{d L}{L} \frac{\partial k(L)}{\partial
L},\label{einsol}\eeq
\noindent
where ${1\over3}V$ is a constant of integration of \myref{eincond}
over $L$, and is therefore independent of $L$.  If we take
\bea V &=& - bV(Z,\Z) + \half\dx V_X,\qquad
V(Z,\Z) = G(T,\T) + V_{\rm inv}(Z,\Z),\label{defV}\eea
\noindent
with $V_{\rm inv}$ modular invariant, under an anomalous
transformation we have $\Del V = H(T,\Lambda_X) +
\H(\T,\bar\Lambda_X)$, with $H$ given by \myref{defH}, and
\beq \Del\L_{\rm K E} =  {1\over8}\superint{E\over8R}\chiproj L H\hc =
-{1\over8}\superint{E\over R}\Phi H\hc,\label{delL2}\eeq
\noindent
since the term involving $\bD^2$ vanishes by partial
integration~\cite{bggm}.  The anomaly \myref{DelL} is canceled:
$\Del\L_{\rm K E} = -\Del\L_{\rm anom}$.

Now consider the following Lagrangian
\beq\L_{\rm lin} = -3\superint \, E \[F(Z,\Z,\vx,L) 
+ \frac{1}{3} (L + \Omega)(S+\bar{S})\],\label{Llin}\eeq
where $S\;(\S)$ is chiral (antichiral):
\beq S = \chiproj\Sigma,\qquad \S = \bchiproj\Sigma^{\dag},\qquad
\Sigma\ne\Sigma^{\dag},\eeq
with $\Sigma$ unconstrained; $L= L^\dag$ is real but otherwise unconstrained,
and $\Omega$ is a real superfield that satisfies

\noindent
\beq (\bar\D^2 - 8R)\Omega = \Phi,\qquad (\D^2 - 8\bar R)\Omega =
\bar\Phi.\eeq
\noindent
If we vary the Lagrangian \myref{Llin} with respect to the
unconstrained superfields $\Sigma,\Sigma^\dag$, we recover the
modified linearity condition \myref{lincond}. This results in the term
proportional to $S+\S$ dropping out from \myref{Llin}, which reduces
to \myref{Llin1}, with
\beq  F(Z,\Z,\vx,L) = 1 - {1\over3}L\[2s(L) - V(Z,\Z,\vx)\],\qquad
s(L) = - \half\int{d L\over L}{\pp k(L)\over\pp L},\eeq
where the vacuum value $\langle{\l s(L)\r}\rangle = \langle{s(\ell)}
\rangle = g^{-2}_s$ is the gauge coupling constant at the string scale.

Alternatively, we can vary the Lagrangian \myref{Llin} with respect to
$L$, which determines $L$ as a function of $S + \S + V$, subject
to the condition
\beq F(Z,\Z,\vx,L) + {1\over3}L(S + \S) = 1\equiv
F(Z,\Z,\vx,S+\S + V),\label{einst}\eeq
\noindent
which assures that once the (modified) linear multiplet is eliminated,
the requirement
$$\L_{\rm K E} = - 3\int d^4\theta E F(Z,\Z,\vx,S+\S) = - 3\int
d^4\theta E, \qquad K = K(Z,\Z,\vx,S+\S),$$
\noindent
for a canonically normalized Einstein term with only chiral matter is
recovered.  Together with the equation of motion\footnote{In \uk\,
  superspace $E$ has an implicit dependence on the K\"ahler potential
  $K$ such that $\pp E/\pp L = - E(\pp K/3\pp L)$. With the
  conventions of \cite{bggm}, $\Omega$ has Weyl weight $w_W(\Omega) =
  - w_W(E) = 2$, so $E\Omega$ is independent of $K$, i.e. of $L$, and
  $\del\L_{\rm lin}/\del L = 0$ together with \myref{einst} gives
  \myref{eincond}.} for $L$, the condition \myref{einst} is equivalent
to the condition \myref{eincond} and the Lagrangian \myref{Llin}
becomes
\beq\L_{\rm lin} = -3\superint\,E - \superint\,E(S + \S)\Omega
= -3\superint\,E + {1\over8}\(\superint{E\over R}S\Phi\hc\).\label{Lchi}\eeq
Since $L = L(S + \S + V)$ is invariant under T-duality and \ux, we
require $\Del S = - H$, so the variation of \myref{Lchi} is again
given by \myref{delL2}.  

For other orbifolds with $T$-dependent threshold corrections, the
conditions \myref{wthconds} are modified somewhat, but the
cancellation of the anomaly \myref{DelL} goes through as above.  In
this case the modular anomaly in \myref{DelL} is partially canceled
by the threshold cancellations, and partially canceled by the
``Green-Schwarz'' term encoded in the terms proportional to
$V(Z,\Z,\vx)$ in \myref{einsol} and \myref{einst}.

\subsection{Gauge coupling unification}
The form of $V(Z,\Z)$ in \myref{defV} is not completely determined by
the requirement of anomaly cancellation, because $V_{\rm inv}(Z,\Z)$
can be any invariant function of the chiral supermultiplets.  For
example, in $\mathbf{\mathbb{Z}_N}$ models with just the three
``diagonal'' K\"ahler moduli introduced in \myref{Ttransf}, under the
$SL(2,{\bf Z})$ subgroup the transformations \myref{tduality} reduce
to \myref{Ttransf} and
\noindent
\beq \Phi^a\to e^{-\sum_i q^a_i F^i}  \qquad F^i = \ln(i c T^i +
d),\qquad \sum_i F^i = F(T),\label{Phitransf}\eeq
and the K\"ahler potential takes the form
\noindent 
\beq K = K(L) + G(T,\T) + \sum_a|\Phi^a|^2e^{\sum_i q^a_i g^i} +
O(\Phi^3), \qquad \sum_i g^i = G(T,\T),\label{Kpot}\eeq
\noindent
with
\noindent
\beq  g^i \to g^i + F^i + \bF^i \eeq
\noindent
under a T-duality transformation.  If we took, for example, 
\noindent
\beq V_{\rm  inv} =  \sum_{a = 1}^n  c_a\ln\(|\Phi^a|^2e^{\sum_i q^a_i
  g^i}\),  \qquad \sum_{a=1}^n c_a  q^a_i =  - 1\quad  \forall\quad i,
\eeq
\noindent
the K\"ahler moduli would drop out of $V(Z,\Z)$, but the anomaly would
still be canceled.  

In fact, the $T$-dependence of $V$ has been determined~\cite{gt,mskl}
by matching string theory calculations to the effective field theory.
For example, in $\mathbf{\mathbb{Z}_3}$ and $\mathbf{\mathbb{Z}_7}$
orbifolds, with no threshold corrections, the $T$-dependence drops out
of the coefficient of $F\cdot\tF$ when the matter fields $\Phi^a$ are
set to zero.  By supersymmetry, which implies a holomorphic gauge
kinetic function, the coefficient of $F\cdot F$ also vanishes, which
means that the contribution from the field theory loop corrections
must be exactly canceled by that from the ``Green-Schwarz'' term in
\myref{einsol} or \myref{einst}; this requires~\cite{gt} $V_{\rm
  inv}(T,\T,\Phi=0) = 0$.  This result remains true for orbifolds with
threshold corrections, but the coefficient $b\to b_{\rm loop}\ne b$ of
the loop correction in~\myref{DelL}, \myref{defH} is modified in such
a way that the full anomaly is canceled in the presence of additional
T-dependent threshold contributions.

In the regulated theory, the coefficient $g_{a\;\rm eff}^{-2}$ of
$F^a\cdot F_a$ at the string scale is determined by the masses of the PV
fields that replace the cut-off $\Lambda$.  The relevant PV masses are
uniquely determined by the requirement of the cancellation of
ultra-violet divergences. For orbifolds with no threshold corrections
one gets
\noindent
\bea {1\over g_{a\;\rm eff}^2} &=& {1\over g^2(\ell_0)}
-{1\over16\pi^2}\(C^a-C^a_M\)k(\ell_0)- \frac{2}{16\pi^2} \sum_b
C^a_b\ln (1-p_b\ell_0)\, \nnn {1\over g^2(\ell)} &\equiv& s(\ell) =
  -\int d\ell{k'\over2\ell},\label{deltal}\eea
\noindent
where $p_b$ is the coefficient of $|\Phi^b|^2 e^{\sum_i q^b_i g^i}$ in
$V(Z,\Z,\vx)$, and $\ell_0 = \myvev{\ell}$ is the vacuum value of the
scalar component $\ell$ of $L$. The expression \myref{deltal} is
independent of the renormalization scale, and may be
compared~\cite{gt} with the two-loop order renormalization group
invariant quantity~\cite{russians}
\noindent
\begin{equation}
\delta_a=\frac{1}{g_a^2(\mu)}-\frac{1}{16\pi^2}
(3C^a-C_M^a)\ln\mu^2+\frac{2C^a}{16\pi^2}\ln g^2_a(\mu) 
+\frac{2}{16\pi^2}\sum_bC_b^a\ln Z_b^a(\mu)\, ,
\label{rge}\end{equation}
\noindent
where $Z_b^a$ are the renormalization factors for the matter fields,
$\mu$ is the renormalization scale.  If we equate the
scale-independent quantity $\delta_a$ with $g_{a\;\rm eff}^{-2}$ and
impose the boundary conditions
\noindent
\begin{equation} g(\ell_0) = g_s = g_a(\mu_s), \qquad 
Z_b^a(\mu_s)=(1-p_b l)^{-1},\qquad k(\ell_0) 
= \ln\mu_s^2,\label{id}\end{equation}
\noindent
where $\mu_s$ is the string scale in reduced Planck mass units: $m_P =
(8\pi G_N)^{-\half} =1$, we obtain the renormalization
  group equation
\noindent
\bea g_a^{-2}(\mu) &=& g^{-2}(\mu_s) - \ep_a - {1\over8\pi^2}\(3C^a -
C^a_M\)\ln(\mu_s/\mu) + {C^a\over8\pi^2}\ln\[{g^2(\mu_s)/g_a^2(\mu)}\]
\mmm + {1\over8\pi^2}\sum_b C^a_b\ln[Z_b(\mu_s)/Z_b(\mu)],\eea
where 
\noindent
\beq \ep_a = {C^a\over8\pi^2}{\ln g^2(\ell_0)\over k(\ell_0)}\eeq
\noindent
is a scale-independent threshold correction. For example, in the
classical limit we have
\noindent
\beq k(\ell) = \ln\ell, \qquad g^{-2}(\ell) = s(\ell) = - \int
d\ell{k'\over2\ell} = {1\over2\ell},\qquad \ep_a =
\ln2{C^a\over8\pi^2}.\eeq
\noindent
This gives for the gauge unification scale in
the $\overline{MS}$ scheme~\cite{kap}
\noindent
\beq  \mu_{\rm unif}^2={\mu^2_s\over2e} = 
{g^2_s m_P^2\over2e}\sim 2\times 10^{17}{\rm GeV}.\eeq
\noindent
This is an order of magnitude lower than what is obtained by
extrapolating from low energy data~\cite{amaldi} in the context of the
minimal supersymmetric extension of the Standard Model, but in
effective theories from superstrings one expects heavy states that are
vector-like under the Standard Model gauge group, as well as
corrections to the dilaton K\"ahler potential from string
nonperturbative effects and/or field theory loop effects.  For
orbifold compactifications with threshold corrections, there are
additional $T$-dependent terms in \myref{deltal}; these give small
corrections in the weak coupling limit $T\sim 1$.

Note that the result~\myref{deltal} of the one-loop calculation
incorporates the two-loop result in~\myref{rge}.  This is because a
supersymmetric regularization procedure necessarily gives a
supersymmetric result~\cite{bmk2}. The chiral anomaly, which
is completely determined at one loop, must form a supersymmetric
operator with the conformal anomaly.  This two-loop ``correction''
to the standard one-loop form of the beta-function is encoded in
the dilaton dependence of the effective cut-offs, in this case
the PV masses.

\subsection{Hidden gaugino condensation} \label{hid}
A popular candidate for supersymmetry breaking in the context of
superstring theory is through gaugino condensation in a hidden sector,
that is, a Yang-Mills sector that couples to the Standard Model only
through gravitational strength couplings.  Effective theories for
gaugino and matter condensates were first constructed in globally
symmetric theories~\cite{venez}, by matching the anomalies of the
effective condensate Lagrangian to those of the underlying Yang-Mills
Lagrangian.  This can be generalized~\cite{bgw} to the supergravity
case by introducing chiral superfields with \uk\, weight 2 and 0,
respectively, for gaugino condensates $U_a$ and matter condensates
$\Pi^\alpha_a$: 
\noindent
\beq U_a\simeq(\Wc_a\Wa^a)_{\rm hid}, \qquad
\Pi^\alpha_a\simeq\prod_b(\Phi_a^b)^{n^\alpha_b}_{\rm
  hid},\label{cond}\eeq
\noindent
where the elementary chiral field $\Phi^b_a$ is charged under the
strongly coupled hidden sector gauge group $\G_a$. The effective
Lagrangian for these fields is
\noindent
\beq \L_{\rm eff}(U_a,\Pi^\alpha_a) = {1\over8}\int d^4\theta{E\over R}
\sum_a U_a\[b'_a\ln\(e^{-K/2}U_a\) + \sum_\alpha
b^\alpha_a\ln\Pi^\alpha_a\]\hc
\label{LeffU}\eeq
\noindent
with the constant coefficients
\noindent
\beq b'_a = {1\over8\pi^2}\(C^a - \sum_b C^a_b\),\qquad
b^\alpha_a = \sum_{b\in\alpha}{C^a_b\over4\pi^2d^\alpha_a},\qquad
d^\alpha_a = {\rm dim}\(\Pi^\alpha_a\),\label{coeffs}\eeq
\noindent
determined~\cite{bgw,mints} by requiring that the variation of
\myref{LeffU} reproduce the variation \myref{DelL}, \myref{defH} of
the underlying theory, with $U_a$ identified as in \myref{cond}, and
by matching the other anomalies of the effective theory to those of
the underlying theory, including the anomalies under \uk\,
(R-symmetry) and conformal transformations.  Since the right hand
side of the modified linearity constraint \myref{lincond} now has
$W^aW_a$ in \myref{Phi} replaced by $U_a$ for the strongly coupled
hidden gauge groups, overall modular and \ux\, invariance are restored
as before.

Adding a gauge invariant superpotential to the effective theory
\noindent
\beq W(\Pi) = \sum_{a,\alpha}C^a_\alpha(T)\Pi^\alpha_a,\label{WPi}\eeq
\noindent
where the T-dependence of the coefficients $C$ assures invariance
under T-duality, leads to a solution to the equations of motion with
nonvanishing condensate vacuum values and masses of order of the
condensate scale or larger.  Integrating out these heavy condensates
gives a potential for the scalar moduli $t^i,\,s$.  This potential
always has a minimum at the vanishing coupling limit $\myvev{s} =
g^{-2}(\mu_s)\to\infty$, with no gaugino condensate and no
supersymmetry breaking.  In fact, this is the only minimum in the
absence of the symmetry-restoring Green-Schwarz term in \myref{einsol}
or \myref{einst} if the classical form $k(L) = \ln L$, or equivalently
$k(S,\S) = -\ln(S + S)$, of the dilaton K\"ahler potential is used.
This was known as the ``runaway dilaton problem''.  However, when the
Green-Schwarz term is included, there is a second runaway direction,
this time in the direction of strong coupling, where string
nonperturbative effects cannot be ignored.  Including these effects
provides a mechanism~\cite{bgw1} for dilaton stabilization, known as
K\"alher stabilization, at finite coupling and with nonvanishing
condensates and supersymmetry breaking.

\subsection{Axion physics}
The last term in the Lagrangian \myref{Lchi}, with $\Phi = \WaWa$ has
a classical R-symmetry, under which the Yang-Mills fields strengths
$\Wa$ and the condensates $U$ transform, respectively, as
\noindent
\beq \Wa^a(\theta)\to W'^a_\alpha(\theta') =
e^{{i\over2}\alpha}\Wa(\theta'),\qquad U_a(\theta)\to U'_a(\theta') =
e^{i\alpha}U(\theta')\label{Rtransf}\eeq
\noindent
where $\alpha$ is a constant parameter, and $\theta'$ is related to
$\theta$ in such a way that the integral over $\theta$ in
\noindent
\beq \int d^4\theta{E'(\theta')\over R'(\theta')}e^{i\alpha}
\Phi(\theta') = \int d^4\theta'{E(\theta')\over R(\theta')}
\Phi(\theta') = \int d^4\theta{E(\theta)\over R(\theta)}\Phi(\theta),
\label{LPhi}\eeq
\noindent
is invariant.\footnote{The integral $\int d^4\theta E/R$ in local
  supersymmetry transforms the same way as $\int d^2\theta \to
  e^{-i\alpha}\int d^2\theta'$ in global supersymmetry.} 
For an arbitrary chiral superfield $\Phi$ such that
\noindent 
\beq \Phi'(\theta') = e^{i\beta}\Phi(\theta'),\label{Phitransf2} \eeq
\noindent 
under R-symmetry with gauge superfields transforming as in
\myref{Rtransf}, the component fields transform as
\noindent
\beq \l{\pp^n\over\pp\theta^n}\Phi(\theta)\r\to e^{i(\beta -
{n\over2}\alpha)}\l{\pp^n\over\pp\theta^n}\Phi(\theta)\r,
\qquad n = 0,1,2.\label{comps}\eeq
\noindent
The symmetry under \myref{Rtransf} is anomalous at the quantum level.
For example, under \uk\, the gauge supermultiplets transform as in
\myref{Rtransf}, while matter chiral supermultiplets transform simply
as
\noindent
\beq \Phi^A_a(\theta) \to \Phi^A_a(\theta'),\qquad 
\Pi^\alpha_a(\theta) \to \Pi^\alpha_a(\theta'),
\label{Pitransf1}\eeq
\noindent
and the shift in the Yang-Mills coupling in \myref{Lchi} is given by
\noindent
\beq \Del\L_{\rm Y M}(\alpha) = {i\alpha\over8}\sum_a b'_a\int d^2\theta
\WaWa \hc = -{\alpha\over4}\sqrt{g}\sum_a b'_a F_a\cdot\tF^a,
\label{DelLR}\eeq
\noindent
with $b'_a$ given in \myref{coeffs}.  In the absence of
nonperturbative effects, the right-hand side of \myref{DelLR} is a
total derivative, and has no effect on the S-matrix.  This is no
longer true when condensation occurs, and $\WaWa$ is replaced by the
condensate $U_a$ for one or more gauge group factors $\G_a$. Then
R-symmetry is generally broken, just as quark condensation in QCD
breaks chiral symmetry.  However, if the Lagrangian is independent of
the axion $a =\im s$ except for its coupling to $\Phi$ in
\myref{Lchi}, there is a residual R-symmetry~\cite{bd} in the case of
just one condensate $U_c$.  The variation of the condensate term in,
e.g., \myref{DelLR} can be compensated by for a shift in the axion:
\noindent
\beq \im s  \to\im s - \alpha b'_c.\label{DelS}\eeq
\noindent

In the class of models for gaugino condensation discussed above, the
``classical'' condensate Lagrangian (i.e. the part without the logs)
is not invariant under \uk\, in the presence of the superpotential
\myref{WPi}, which gives a superpotential Lagrangian term of the form
\myref{LPhi} with $\Phi = \half e^K W(\Pi)$.  In this case the
classical R-symmetry has, instead of \myref{Pitransf1}
\noindent
\beq \Phi^{b\in\alpha}_a(\theta) \to e^{i\alpha/d^\alpha_a}
\Phi^{b\in\alpha}_a(\theta'),\qquad \Pi^\alpha_a(\theta) \to 
e^{i\alpha}\Pi^\alpha_a(\theta'),
\label{Pitransf2}\eeq
\noindent
and, for the strongly coupled condensate $U_c$, the shifts \myref{DelLR},
\myref{DelS} are replaced by
\noindent
\beq \Del\L_{\rm Y M}(\alpha) = {i\alpha\over8}b''_c\int d^2\theta U_c
\hc, \qquad \Del\im s = - \alpha b''_c,\qquad b''_c = b'_c +
\sum_{\alpha}b^\alpha_c. \label{DelLR2}\eeq
\noindent
If the condensates $\Pi^\alpha_c$ have dimension 3, from
\myref{coeffs} we have simply
\noindent
\beq b''_c = b_c = {1\over8\pi^2}\(C^c - {1\over3}\sum_A C^c_A\),
\label{beta}\eeq
\noindent
which is related to the $\beta$-function by the one-loop order RGE
\noindent
\beq {\pp g_a(\mu)\over\pp\ln\mu} = - {3b_a\over2}g_a^3(\mu).\eeq
\noindent
The dilaton potential\footnote{In K\"ahler stabilization models the
  T-moduli are generically stabilized at self-dual points with
  vanishing F-terms, and supersymmetry breaking is dilaton dominated.
We fix $t^i = \l T^i\r$ at these self-dual points in what follows.}
is dominated by the gauge group $\G_c$ with the largest
beta-function coefficient $b_c$ and the largest condensation scale
\noindent
\beq \Lambda_c \sim e^{-1/3b_cg^2_s}\mu_s, \qquad \myvev{|u_c|}
\sim \Lambda_c^3, \qquad u_c = \l U_c\r.\eeq
\noindent
The dilaton acquires a mass of order $\myvev{|u_c|}$ in reduced Planck
units, but the axion remains massless if there is a single condensate.
In this case it is a prime candidate for the QCD axion.

If there is more than one term in the sum over $a$ in \myref{LeffU},
the axion acquires a small mass $m_a$.  For example if there are two
strongly coupled gauge groups $\G_c,\,\G_d$, both with
dim$\(\Pi^\alpha_{c,d}\) = 3$, we get~\cite{bgw}
\beq m_a \sim (b_c -b_d)\myvev{\sqrt{|u_1u_2|}}.\eeq
\noindent
This two-condensate system has a point of enhanced symmetry where the
$\beta$-functions are equal, and R-symmetry remains unbroken.  If the
string $a$ axion is to play the role of the QCD, or Peccei-Quinn,
axion its mass due to symmetry-breaking must be much smaller than the
QCD condensation scale.  The only realistic possibility in the
heterotic string context is that of a single hidden sector gaugino
condensate.\footnote{A possibly dangerous contribution to the axion
  mass is from higher dimension operators~\cite{bd} such as $\L\ni
  \sum_n\int d^4\theta(E/R)c_n(Z)U_c^{n+1}m_P^{-3n}$.  However, the
  dimension of these operators is severely restricted by
  T-duality~\cite{butter}.  The minimal T-duality $SL(2,\bf Z)$ group
  of \myref{Ttransf} requires $n\ge 4$ which gives a contribution that
  could be comparable to the axion mass generated by the QCD
  condensate~\cite{gk} if $\myvev{|u_c|}/m^3_P\sim 10^{-12}$, and
  therefore problematic.  However any group larger than the minimal
  one should result in a negligible contribution.  For example
  $SL[(2,\bf Z)]^2$ and $SL[(2,\bf Z)]^3$ require $n\ge 8$ and $n\ge
  12$, respectively.}

The Peccei-Quinn symmetry was introduced to eliminate the $CP$
violating term in the QCD Lagrangian
\noindent
\beq \L_{\rm QCD} = {\theta\over32\pi^2}F^{a}\cdot\tF_a,\label{theta}
\eeq
\noindent
where $F^a_{\mu\nu}$ is the gluon field strength, that is expected to
contribute to the S-matrix in the presence of nonperturbative strong
coupling effects. The term in \myref{theta} can be rotated away by a
chiral transformation on quarks because the associated anomaly
generates a term of the same form.  However, unless there is at least
one massless quark, any chiral symmetry is broken by quark masses, and
CP violation reappears in the form of phases in the quark mass matrix.
CP conservation is preserved only if there is a nonamalous symmetry
involving chiral transformations on the quarks such that that $\theta$
in \myref{theta} can be set to zero in the basis in which the quark
mass matrix is real.  If this symmetry is broken only by quark masses
much smaller than the QCD condensation scale, there will be a small
violation of CP that is acceptable as long as $\bar\theta$, the value
of $\theta$ in the real quark mass basis, is less than $10^{-10}$ as
required by the stringent limits on the neutron dipole moment.

A convenient toy model for studying the axion mass is supersymmetric
$SU(N_c)$ with $N$ flavors, i.e. $N$ quark and $N$ antiquark chiral
superfields $Q^A,\,Q^c_A$.  The effective theory for a condensate in
this case can be constructed as above, except that the matter
condensate 
\noindent
\beq \Pi^\alpha_Q = \det{\bf\Pi},\qquad {\bf\Pi}^A_B = Q^A
Q^c_B,\qquad \dim\Pi^\alpha_Q = 2N,
\label{PiSUN}\eeq
\noindent
is determined by the requirement of invariance under the nonanomalous
symmetry $SU(N)_L\otimes SU(N)_R$.  The condensate Lagrangian takes
the form
\noindent 
\bea \L(U_Q) &=& {1\over8}\int d^4\theta{E\over R}U_Q\[S + b'_Q\ln U_Q
+ b^\alpha_Q\ln\det{\bf\Pi}\]\hc,\nnn b'_Q &=& {1\over8\pi^2}\(N_c - N\),
\qquad b^\alpha_Q = {1\over8\pi^2}.\label{LUQ}\eea
\noindent
If we add a superpotential for $\bf\Pi$,
\noindent
\beq W({\bf\Pi}) = \Tr\[{\bf C}(T){\bf\Pi C^c}(T){\bf M}\],\label{WPi2}
\eeq
\noindent
where ${\bf M}$ is the quark mass matrix and $\bf C,\,C^c$ are
matrix-valued functions of the K\"ahler moduli that assure T-duality
invariance of the ``classical'' condensate Lagrangian.  The classical
Lagrangian is also invariant under an R-symmetry if $U_Q$ transforms
as in \myref{Rtransf} and
\noindent
\beq {\bf\Pi}\to e^{i\alpha}{\bf\Pi},\qquad \det{\bf\Pi}\to
e^{iN\alpha}\det\bf\Pi.\label{Pitransf3}\eeq
Then \myref{LUQ} transforms as 
\noindent
\beq \Del\L(U_Q) = {1\over8}\int d^4\theta{E\over R}U_Q\[\Del S
+ i\alpha\(b'_Q + N b^\alpha_Q\)\]\hc,\label{DelLUQ}\eeq
\noindent
and is invariant provided
\noindent
\beq \Del S = i\Del\im s = -i\alpha\(b'_Q + N b^\alpha_Q\)
= - i{\alpha N_c\over8\pi^2}.\eeq
\noindent
In the global supersymmetry limit, $m_P\to\infty$, $\re s\to g^{-2}$,
the results found for supersymmetric Yang-Mills theories using
holomorphic arguments~\cite{davis} are recovered in this effective
theory~\cite{gk}.

If a confined sector with dimension-three matter condensates is also
present, there is a nonanomalous R-symmetry in the absence of quark
masses.  It is defined by \myref{Rtransf},
\myref{Pitransf2}-\myref{beta} and 
\noindent
\beq {\bf\Pi}\to e^{i\beta}{\bf\Pi},\qquad \det{\bf\Pi}\to
e^{iN\beta}\det{\bf\Pi},\qquad \beta = \alpha\({8\pi^2b_c - N_c\over N}
 + 1\).\label{Pitransf4}\eeq
\noindent
In the presence of quark masses, $\det{\bf M}\ne 0$, the R-symmetry
is broken, except at the point of enhanced symmetry:
\noindent
\beq \beta = \alpha,\qquad b_c = {N_c\over8\pi^2},\label{sympt}\eeq
\noindent
and the axion acquires a mass~\cite{gk}
\noindent
\beq m_a = {F_\pi\over F_a}{|8\pi^2b_c - N_c|\over\sqrt{2n}b_c}m_\pi,
\label{ma}\eeq
\noindent
where $n$ is the number of flavors with quark masses below the QCD
condensation scale (here taken to be degenerate), $m_\pi$ is the 
common mass of the corresponding light pseudoscalars, $F_\pi$ is 
the pion decay constant (93 MeV in QCD), 
\noindent  
\beq a = \myvev{\sqrt{2\ell/k'(\ell)}}\im s = F_a\im s\eeq
\noindent
is the canonically normalized axion, and $F_a$ is its coupling to
the Yang-Mills sector at the string scale in reduced Planck units:
\noindent 
\beq \L_{\rm st}\ni - {\im s\over4}\sum_a F^a\cdot\tF_a = - {a\over4F_a}
\sum_a F^a\cdot\tF_a.\label{Lst}\eeq

In the case that we are actually interested in, QCD condensation
occurs well below the scale of supersymmetry breaking, and one must
find the correct effective pion-axion theory by first integrating out
the heavy superpartners of Standard Model particles, as well as the
heavy quarks.  However the result in \myref{ma} is essentially
unchanged; for just two light quarks $u,d$, it is simply multiplied by
a function of the quark mass ratio:
\noindent
\beq \l m_a\r_{n=2} \to {2\sqrt{z}\over 1+z}\l m_a\r_{n=2},
\qquad z = {m_u\over m_d}.\eeq
\noindent
The result \myref{ma} appears to be at odds with the well-known
relation~\cite{sred} between the axion mass and its coupling strength.
However $F_a$ is the axion coupling at the {\it string scale}.  When
the gauginos are integrated out, their coupling to the axion generates
new terms in the couplings of the axion to gauge field strengths.  For
the axion coupling to QCD gluons, one gets a contribution~\cite{gk}
\noindent
\beq \Del\L_{\rm QCD} = {a\over4 F_a}{N_c\over8\pi^2}(F\cdot\tF)_Q.\eeq
\noindent
Combining this term with the QCD term in \myref{Lst} one gets
for the total axion-gluon coupling at low energy
\noindent
\beq \L_{\rm QCD} \ni - {a\over4F_a}\(1 - {N_c\over8\pi^2}\)(F\cdot\tF)_Q
\equiv - {na\over32\pi^2f_a}(F\cdot\tF)_Q,\eeq
\noindent 
where we have introduced an alternative normalization $f_a$ for the
axion coupling that is often found in the literature.  In terms
of this parameter, the axion mass for $n=2$ takes the familiar form
\beq m_a = {2\sqrt{z}\over1 + z}{F_\pi\over f_a}m_\pi.\eeq
\noindent
The axion mass vanishes at the point of enhanced symmetry
\myref{sympt} and one loses a potential solution to the ``strong CP
problem''.  This is because, under \myref{Phitransf2}, \myref{comps}
and \myref{Pitransf4} the quark superfields $Q$ transform with phase
$\half\beta$ and the quarks $q = \l\pp Q/\pp\theta\r$ with phase
\noindent
\beq \half\(\beta - \alpha\) = \alpha{8\pi^2b_c -
  N_c\over2N},\eeq
\noindent
which vanishes at the symmetry point, so the nonanomalous R-symmetry
does not affect the quark mass matrix and cannot be used to set
$\theta$ to zero in the basis where the masses are real.

\section{Soft supersymmetry breaking at one-loop}
When supersymmetry is broken in a hidden sector of a generic
supergravity theory, the Lagrangian for the ``observable''
(i.e. Standard Model) sector acquires ``soft'' supersymmetry-breaking
terms.  These are terms of dimension two or three that do not affect
the cancellations of ultraviolet divergences that are present in the
supersymmetric theory.  They include gaugino masses, holomorphic
functions of chiral scalars that are cubic (A-terms) and quadratic
(B-terms), and ``soft'' scalar masses, that, is scalar squared mass
terms $m^2_{\bj i}\bph^{\bj}\phi^i$ that have no fermionic
counterparts. However, there are cases where these terms are absent at
tree level.  For example, if supersymmetry is F-term mediated,
$\myvev{D_a}=0,\;\myvev{F^T}\ne0$, the so-called ``no-scale'' K\"ahler
potential of \myref{noscale} leads to vanishing soft terms at tree
level if the superpotential is independent of the K\"ahler moduli
$T^i$.  In such cases loop-induced soft terms become important, as
they do if some tree-level soft terms are suppressed.  For example, in
the class of models outlined in \mysec{hid}, the gaugino masses and
A-terms are much smaller than soft scalar masses~\cite{bgw2} if the
$\beta$-function coefficient $b_c$ of the dominant condensing gauge
group is an order of magnitude or so smaller than the parameter $b$
appearing in the Green-Schwarz term \myref{defV}.  

The loop corrections to soft supersymmetry-breaking terms include
the so-called ``anomaly mediated'' contributions that are present even
when there are no soft terms at tree-level; they arise from the
super-Weyl anomaly of standard supergravity.  Some of these are model
independent in the sense that they are determined only by the
$\beta$- and $\gamma$-functions appearing in the RGE's of the low
energy theory, and are independent of the mechanism by which
supergravity is broken. That is, they are independent of the vacuum
values of the auxiliary fields except for the the supergravity
auxiliary field, whose vacuum value $\myvev{R} = M_\psi$ signals the
breaking of {\it local} supersymmetry.

The ``model-independent'' contribution to gaugino masses $m_a$ was
first identified in~\cite{rs,glmr}.  This result was subsequently
confirmed and completed~\cite{gnw,bagger,gn}, with the result
\noindent
\beq m_a^{\rm 1-loop} = - {g^2_a(m_a)}\({3\over2}b_a M_\psi + \half b'_a
\myvev{F^i K_i} + {1\over8\pi^2}\sum_b C^b_a 
\myvev{F^i\pp_i\ln K_{b\bar b}}\),\label{mg}\eeq
\noindent
where the first term is the ``model independent'' one alluded to
above.  In the models of interest here, only the auxiliary fields
$F^S,\,F^{T^i}$ have nonvanishing vacuum values at the hidden sector
supersymmetry scale, and the class of condensation models described in
\mysec{hid} generally have $\myvev{F^{T^i}} = 0$.  When the 4d
Green-Schwarz term and threshold effects are included, there are
additional contributions to the gaugino masses; these modify only the
coefficients of $\myvev{F^{T^i}}$.  The full expressions for the
gaugino masses as well as other soft terms are given in~\cite{bgn}. 

The result \myref{mg} was obtained by analyses~\cite{gnw,bagger} of
the loop-induced operator that transforms as
in \myref{DelL}, by using~\cite{pom} spurion techniques~\cite{grisaru},
and by an explicit PV calculation~\cite{gnw}.  In the last case, the
``anomaly-induced'' gaugino mass results from a B-term insertion on
the squark lines in PV squark-quark ($\phi^P$-$\chi^P$) loop
contributions to the gaugino masses:
\noindent 
\beq \L_{\rm P V}\(\phi^P\)\ni - e^KW_{\rm P V}(\phi)\myvev{\overline
  W}\hc = - e^K\mu_{P Q}\phi^P\phi^Q\M_\psi\hc\eeq
\noindent

Model independent independent contributions to $A$-terms were also
found using spurion techniques~\cite{pom,grisaru}, namely:
\noindent
\beq A^{\rm 1-loop}_{a b c} \ni \(\gamma_a + \gamma_b + \gamma_c\)M_\psi,
\label{Aterm}\eeq
\noindent
where $\gamma_a$ is the anomalous dimension for the light
supermultiplet $\phi^a,\chi^a$.  However, this technique failed to
yield an analogous contribution to soft scalar masses, for which a
such a contribution appeared only at two-loop order, proportional to
the derivative of the anomalous dimension.

Pauli-Villars calculations~\cite{gn} of these effects confirmed the
contribution \myref{Aterm}, and a similar B-term contribution
\noindent
\beq B^{\rm 1-loop}_{a b} \ni \(\gamma_a + \gamma_b\)M_\psi,
\label{Bterm}\eeq
\noindent
but in addition yielded a one-loop model-independent
contribution to soft scalar masses:
\noindent
\beq (m_a^2)^{\rm 1-loop}_{\rm soft} \ni \gamma_a|M_\psi|^2.
\label{softm}\eeq
\noindent
The source of this discrepancy can be traced to the fact that in the
earlier calculations a holomorphic form for the supersymmetry-breaking
spurion was assumed.  This corresponds, in PV language, to PV B-terms,
but no soft PV masses.  If only B-terms were present soft scalar
masses could result from a double B-term insertion, but these
contributions in fact cancel; the result in \myref{softm} instead
arises from a PV soft (squared) mass insertion.  Repeating the spurion
analysis without the assumption of holomorphism indeed
reproduces~\cite{gn} the term in \myref{softm}.

The compete expressions for the soft supersymmetry breaking terms in
the scalar potential are quite complicated.  They depend on the
tree-level soft terms, as well as on unknown mass parameters in the
Pauli-Villars sector.  This is in contrast to the result \myref{mg},
which is completely determined by the low energy theory.  The PV
masses depend on the PV K\"ahler metric.  In the case of gauge
couplings, {\it all} PV chiral multiplets that are charged under
$\G_a$ contribute both to the ultraviolet divergences associated with
the loop-induced Yang-Mills operator containing $\WaWa$ and to the
soft masses $m_a$; their gauge-charge weighted masses are constrained
by the requirement of ultraviolet finiteness.  On the other hand, only
a subset $\Phi^P$ of PV chiral multiplets contribute to the
renormalization of the K\"ahler potential through their couplings to
light fields $\phi^p$ in superpotential terms $W \sim
\Phi^P\Phi^Q\phi^r$. The K\"ahler metric of the $\Phi^P$ is fixed by
the finiteness requirement. However each PV field $\Phi^P$ has a PV
mass coupling to some other field $\Phi^{P'}$ which need have no
coupling to light fields and no restriction on its K\"ahler metric.
As a result the masses
\noindent
\beq m_P = e^{K/2}(K_{P P}K_{P' P'})^{-\half}\mu_{P P'}\eeq
of the fields $\Phi^P$ that contribute to the scalar soft terms are
not fixed by finiteness alone, and depend on the details of
string/Planck scale physics.

\section{Unfinished business}\label{unfin}
The F-term anomalies of the form \myref{DelL} that are linear in the
parameters of the anomalous transformations are well understood and
can be canceled by a combination of the 4d Green-Schwarz term and
string loop threshold effects.  However the anomalous terms that are
higher order in these parameters depend on the details of the
regularization procedure~\cite{bg,ssanom}.  It appears likely that if
a regularization procedure can be found that allows for the
implementation of full anomaly cancellation in the context of the
weakly coupled heterotic string, it will entail some constraints on
higher order terms in the K\"ahler potential~\myref{Kpot} for the
untwisted sector fields $\Phi^a$ of orbifold compactifications.  This
is reminiscent of the constraint \myref{cascond} on gauge charges, and
could have important implications for flavor changing neutral currents
discussed in \mysec{fcnc}.  Such a procedure would certainly entail
restrictions on the PV masses, which, as discussed in the previous
section, play a role in soft supersymmetry breaking parameters in the
case that loop corrections to these are important. Therefore the
solution to the problem of full anomaly cancellation will have direct
implications for phenomenology.

Finally, as mentioned \mysec{PV} there are also D-term anomalies, and
as yet there has been no serious attempt to determine how they might
be canceled in a string theory context.  The resolution of this issue
may also have a bearing on the effective supergravity Lagrangian.

\section{Final word}
Raymond Stora was a cherished friend and colleague who was always very
supportive.  I had the great pleasure of organizing with Raymond a
very successful summer school at Les Houches in 1981; most of the
participants in that session are still active in particle physics
today. Bruno Zumino and I shared many pleasurable occasions with
Raymond and Marie-France. I will miss him.

\vskip .3in
\noindent{\bf Acknowledgments.} This work was supported in part by
the Director, Office of Science, Office of High Energy and Nuclear
Physics, Division of High Energy Physics, of the U.S. Department of
Energy under Contract DE-AC02-05CH11231, in part by the National
Science Foundation under grants PHY-1316783.

\end{document}